
\input phyzzx

\nopagenumbers
\line{\hfil CU-TP-619}
\line{\hfil gr-qc/9312018}
\vglue .4in
\centerline {\twelvebf  Time-Dependent Open String Solutions in
	2+1 Dimensional Gravity  }
\vskip .3in
\centerline{\it Hai Ren }
\vskip .1in
\centerline{Physics Department}
\centerline{Columbia University }
\centerline{New York, New York 10027}
\vskip .4in
\baselineskip=20pt
\overfullrule=0pt
\centerline {\bf Abstract}
\medskip

	We find general, time-dependent solutions produced by
open string sources carrying no momentum flow
in 2+1 dimensional gravity. The local Poincar\'e group elements
associated
with these solutions and the coordinate transformations that
transform these solutions into Minkowski metric are obtained.
We also find the
relation between these solutions and the planar wall solutions
in 3+1 dimensions.

\vskip 1.1in
\noindent\footnote{}{\twelvepoint This work was supported in part by the US
Department of Energy }

\vfill
\eject

\baselineskip=19pt
\pagenumbers
\pageno=1

\def\refmark#1{[#1]}
\def\onee{\pi G\sigma}
\def\twoo{\pi G\xi}
\def\ga{\alpha}
\def\gb{\beta}
\def\gd{\delta}
\def\gl{\lambda}
\def\ge{\epsilon}
\def\p{\partial}
\def\gG{\Gamma}
\def\w{\omega}
\def\bw{{\w\kern -.60em \w}}
\def\gL{\Lambda}
\def\pz{\partial_z}

\def\pt{\partial_t}
\def\ac{{\pz A\over C}}
\def\ca{{\pt C\over A}}
\def\bc{{\pz B\over C}}
\def\ba{{\pt B\over A}}
\def\tilde{\widetilde}
\def\tt{{\tilde t}}
\def\tz{{\tilde z}}
\def\tx{{\tilde x}}
\def\tonee{\pi G\tilde\sigma}
\def\ttwoo{\pi G\tilde\xi}
\def\tsigma{\tilde \sigma}
\def\txi{\tilde \xi}

\chapter{ Introduction }

	There has recently been much interest in $2+1$ dimensional
gravity \Ref\review{For a review, see R.Jackiw,
{\it Five Lectures on Planar Gravity},
in Cocoyoc 1990, Proceedings, Relativity and gravitation: Classical
and quantum;  {\it Update on Planar Gravity}
({\it Physcis of Infinite Cosmic Strings}),
MIT-CTP-1986, June 1991 (unpublished).}.
Three dimensional Einstein gravity has the important feature
that spacetime is locally flat away from sources. Nevertheless solutions
can be globally non-trivial. It also admits a gauge formulation with the
action given by the Chern-Simons action for the $2+1$ dimensional
Poincar\'e group $ISO(2,1)$ \Ref\CS{A.Achucarro and P.Townsend,
Phys.Lett.{\bf 180}B, 89 (1986);
E.Witten, Nucl.Phys.B{\bf 311}, 46 (1988).}.
In this work we study general, time-dependent solutions
produced by open string sources (which are translational invariant along
the direction of the string) carrying no momentum flow.
Static string solutions
in $2+1$ dimensions were previously studied by Deser and Jackiw
\Ref\Deser{S.Deser and R.Jackiw, Ann.Phys.(NY){\bf 192}, 352 (1989).}.
Stationary string solutions were also studied
\Ref\station{G.Grignani and C.Lee, Ann.Phys.(NY){\bf 196}, 386 (1989);
G.Cl\'ement, {\it ibid.}{\bf 201}, 241 (1990).}.

	The remainder of this paper is organized as follows.
In Sec.II we obtain vacuum solutions away from the string,
and study their gauge formulation.
In Sec.III we
match these vacuum solutions across the string using
junction conditions. We then obtain the relation between
these solutions and the $3+1$ dimensional planar wall solutions
studied by Vilenkin, Ipser and Sikivie \Ref\Vilenkin{A.Vilenkin,
Phys.Lett.{\bf 133}B, 177 (1983).} \Ref\Ipser{J.Ipser and P.Sikivie,
Phys.Rev.D{\bf 30}, 712 (1984).}. Brief conclusions are given in
Sec.IV. We study the geodesic motion of test particles in a typical
background metric in the appendix.

\chapter{ Vacuum Solutions and Their Gauge Formulation}

	In this section we shall find vacuum solutions for the following
``line symmetric'' metric
$$ ds^2 = -A^2(t,z)dt^2 + C^2(t,z)dz^2 + B^2(t,z)dx^2
		\eqn\si  $$
This is the most general form one can have if one requires the metric
be invariant under $x$ translation $x\rightarrow x+a$ and inversion
$x\rightarrow -x$, the $x$-axis being where the
string resides. One can further make $A=C$ by appropriate coordinate
transformations. (It is well known in elementary differential
geometry that any two-dimensional metric can be brought into this
``isothermal'' form.) The position of the string, $z=0$, will in general
be transformed into $z=h(t)$. One can then transform it back to
$z=0$, while retaining the $A=C$ form of the metric, by a transformation
of the form
$$\eqalign{ t+z &\longrightarrow f(t+z) \cr
t-z &\longrightarrow g(t-z) }
	\eqn\eq   $$
where $f$ and $g$ are subject to the constraint
$$ f'(t+z)g'(t-z) > 0
	\eqn\cons  $$
and a prime denotes differentiation of a function with respect to
its argument. Therefore to find general solutions, we only need to
find solutions with $A=C$. However, we would like to leave $A$ and
$C$ arbitrary as far as we can, so that results are directly applicable
to solutions not in the $A=C$ form. It is also {\it easier} to obtain
some particular solutions when $A$ and $C$ are arbitrary. We will
illustrate this later.

	To proceed, we choose the {\it dreibein}
$ {\bf e}_a\equiv e_{a \mu }dx^\mu$ to be
$$ {\bf e}_0 = -A dt , \qquad {\bf e}_1 = C dz , \qquad
{\bf e}_2 = B dx
	\eqn\eq   $$
Latin indices are raised and lowered by
$\eta_{ab} = {\rm diag}(-1,1,1)$ and Greek indices by $g_{\mu\nu}$.
It follows that
the connection $1$-forms $\bw_{ab}\equiv\w_{\mu ab}dx^\mu$ and
the curvature two-forms ${\bf R}_{ab}\equiv R_{\mu\nu ab}dx^\mu\wedge
dx^\nu$ are given by
$$\eqalign{ \bw_{01} &= -\ac dt -\ca dz \cr
\bw_{02} &= -\ba dx    \cr
\bw_{12} &= -\bc dx   }
	\eqn\conn  $$
$$\eqalign{{\bf R}_{01} &=\left[\pz\left(\ac\right) -
\pt\left(\ca\right)\right] dt\wedge dz  \cr
{\bf R}_{02} &=\left[\ac\bc - \pt\left(\ba\right)\right]dt\wedge dx
+ \left[\ca\bc - \pz\left(\ba\right)\right]dz\wedge dx  \cr
{\bf R}_{12} &=\left[\ac\ba - \pt\left(\bc\right)\right]dt\wedge dx
+ \left[\ca\ba - \pz\left(\bc\right)\right]dz\wedge dx }
	\eqn\siv   $$
In $2+1$ dimensions, the curvature tensor is
proportional to the Einstein tensor
$$ R^{\ga\mu}_{\gb\nu} = \ge^{\ga\mu\gl}\ge_{\gb\nu\gd}G^\gd_\gl
	\eqn\prop   $$
Therefore vacuum Einstein equations can be solved by setting
the curvature tensor $R_{\mu\nu\ga\gb}$ to zero. In the case of
static solutions, one has
$$ \pz\left(\ac\right) = 0\ , \qquad\pz\left(\bc\right) = 0\ ,
\qquad (\pz A)(\pz B) = 0
	\eqn\static  $$
To find time-dependent solutions, we define
$$ F \equiv \ac , \qquad G \equiv \ca , \qquad P \equiv \ba ,
\qquad Q \equiv \bc
	\eqn\eq  $$
We then obtain the following five equations. Only four of them
are independent, but all are needed to make the whole set of equations
symmetric.
$$ \eqalign{\pz F &- \pt G = 0 \cr
\pt P &= FQ , \qquad \pt Q = FP  \cr
\pz P &= GQ , \qquad \pz Q = GP  }
	\eqn\eq  $$
The last four equations tell us that $P^2 - Q^2$ is a $(t,z)$ independent
constant. Thus solutions fall into two classes:

\noindent{\bf Type-I solutions}

	The solutions are characterized by
$P^2-Q^2=\pm\gb^2\not=0$. With the $+$ sign, the solutions are given by
$$\eqalign{ \ac &= \pt S , \qquad \ca = \pz S  \cr
 \ba &= \gb\cosh S  , \qquad \bc = \gb\sinh S }
	\eqn\sviii  $$
where $\gb$ is an arbitrary constant and $S = S(t,z)$ an arbitrary
function. Interchanging $\cosh S$ and $\sinh S$ in the above also
gives solutions. This corresponds to choosing the $-$ sign. We will
refer to these solutions as type-I($\pm$) solutions. For simplicity,
the distinction $(\pm)$ will sometimes not be made.

\noindent{\bf Type-II solutions}

	This case corresponds to $P^2 - Q^2 = 0$. The solutions are
$$\eqalign{\ac &= \pm {\pt P\over P} , \qquad \ca = \pm{\pz P\over P} \cr
\ba &= P , \qquad \bc = \pm P  }
	\eqn\six   $$
where $P = P(t,z)$ is an arbitrary function. We choose either all upper
signs or all lower signs.
Eq.\sviii\ with $\gb$ set to zero also gives type-II solutions.

	According to Eq.\prop, the vacuum spacetime is maximally
symmetric with zero curvature. Therefore despite its appearance, the
spacetime is actually homogeneous and isotropic about all points.
The metric can always be transformed into the Minkowski one
$$ ds^2 = -d\tilde t^2 + d\tilde z^2 + d\tilde x^2
	\eqn\Mink   $$
We assume that this coordinate transformation is described by
$$ \p_\mu q^a \equiv {\p q^a\over \p x^\mu}
= \gL^a_{\ b} e^b_{\ \mu}
	\eqn\trans  $$
where $x^\mu\equiv (t,z,x)^{\rm T}$ and
$q^a\equiv (\tt,\tz,\tx)^{\rm T}$. This defines the
(spacetime dependent)
matrix $\gL$. That $\gL$ is a Lorentz matrix follows from the fact
that $e^a_{\ \mu}$ is the {\it dreibein}.
\def\tV{\tilde V}
To see this, consider a vector $V^\mu$ and
its transformation into Minkowski coordinates $\tV^a$.
We have
$$ (\p_\mu q^a)V^\mu = \tV^a
	\eqn\eq  $$
or
$$ V_\mu = e_{b\mu}(\gL^{-1})^b_{\ a}\tV^a
	\eqn\eq  $$
Comparing this with
$$ V_\mu = (\p_\mu q^a)\tV_a = \gL^a_{\ b}e^b_{\ \mu}\tV_a
	\eqn\eq  $$
one obtains in matrix notation
$$ \gL^{-1} = \eta\gL^{{\rm T}}\eta
	\eqn\eq  $$

	There are six Killing vectors. In Minkowski coordinates $q^a$,
they are given by
$$ l^a = c^a
	\eqn\eq  $$
and
$$ l^a = J^a_{\ b}q^b
	\eqn\kill   $$
where $c^a$ are constant vectors and $J$ the Lorentz generators.
If we define the Lie bracket of two vectors $l^a$ and $m^a$ to be
$$ [l^a , m^a] \equiv l^a\p_a m^b - m^a\p_a l^b \equiv n^b
	\eqn\eq   $$
then the Lie bracketing of these Killing vectors
reproduces the commutators
of the Poincar\'e group \Ref\current{See for example, R.Jackiw in
{\it Current algebra and anomalies} by S.B.Treiman, R.Jackiw, B.Zumion
and E.Witten, Princeton University Press (1985).}.
Transforming these vectors into $x^\mu$
coordinates we obtain the vectors
$$ \chi_\mu = \gL^a_{\ b}e^b_{\ \mu}l_a
	\eqn\vec   $$
which must satisfy the Killing equation
$$ \nabla_\mu\chi_\nu + \nabla_\nu\chi_\mu = 0
	\eqn\eq   $$
Substituting Eq.\vec\ into this we obtain, with $l_a$ ``stripped
off'',
\def\d{\nabla}
\def\gLL{\gL^a_{\ b}}
$$\eqalign{ 0 &= \d_\mu(\gL^a_{\ b}e^b_{\ \nu}) +
\d_\nu(\gL^a_{\ b}e^b_{\ \mu})   \cr
&= (\p_\mu \gLL)e^b_{\ \nu} + (\p_\nu \gLL)e^b_{\ \mu}
+ \gLL(\d_\mu e^b_{\ \nu} + \d_\nu e^b_{\ \mu})          }
	\eqn\eq   $$
(The extra contributions obtained when the differential operators act
on the $l_a$ given by Eq.\kill\ cancel due to Eq.\trans\ and the
antisymmetry of the Lorentz generators $J^a_{\ b}$.)
Since
$$ \d_\mu e^b_{\ \nu} = -\w_{\mu\ a}^{\ b}e^a_{\ \nu}
	\eqn\eq  $$
we obtain
$$ (\gL^{-1}\p_\mu\gL)^a_{\ b}e^b_{\ \nu} +
(\gL^{-1}\p_\nu\gL)^a_{\ b}e^b_{\ \mu} = \w_{\mu\ b}^{\ a}e^b_{\ \nu}
+ \w_{\nu\ b}^{\ a}e^b_{\ \mu}
	\eqn\five   $$
Note that, as we will show later,
$(\gL^{-1}\p_\mu\gL)^a_{\ b}$ is anti-symmetric in $a$ and $b$
for {\it any} Lorentz matrix $\gL$.
Eq.\five\ determines
$$ (\gL^{-1}\p_\mu\gL)^a_{\ b} = \w_{\mu\ b}^{\ a}
	\eqn\repro   $$
Eqs.\trans\ and \repro\ reproduce in an instructive way the familiar
equations in the Poincar\'e gauge formulation of the Einstein
gravity. These can also be viewed as a general proof of Gerbert's
claim that the vector $q^a$ satisfying Eq.\trans\
is the local coordinate transformation
that transforms the metric $g_{\mu\nu}$ into $\eta_{ab}$, which he
made based on the study of a class of point source solutions
\Ref\Gerbert{P.Gerbert, Nucl.Phys.B{\bf 346}, 440 (1990).}.

	In terms of the gauge formulation,
since the gauge fields vanish (The connection
$1$-forms Eq.\conn\ already satisfy the torsion free condition,
while Eq.\siv\ sets the Riemann curvature to zero.), the gauge potentials
associated with these vacuum solutions must be pure gauges
$$ A_\mu\equiv e^a_{\ \mu} P_a + \w^a_{\ \mu} J_a = U\p_\mu U^{-1}
	\eqn\sx  $$
Here $U$ is an element of the three dimensional Poincar\'e group
$ISO(2,1)$,
$\w^a_{\ \mu} = -{1\over 2}\ge^{abc}\w_{\mu bc}$,
and $J_a$, $P_a$ are group generators obeying
$$ [J^a,J^b] = -\ge^{abc}J_c \ , \ [J^a,P^b] = -\ge^{abc}P_c\ , \
[P^a,P^b] = 0
	\eqn\eq  $$
with $\ge^{012} = 1$.
We wish to present the Poincar\'e group elements $U$ corresponding
to the solutions Eqs.\sviii\ and \six. In the meantime,
we will find the coordinate transformations that
transform these solutions into the Minkowski metric.
We use a $4\times 4$ representation for $ISO(2,1)$ where an arbitrary
group element $V$ has the form \Ref\group{See for example,
M.Hamermesh,
{\it Group theory and its application to physical
problems}, Addison-Wesley (1962), page 479.}
$$ V = \left(\matrix{\gL^a_{\ b}&q^a\cr 0&1}\right)
	\eqn\eq  $$
where $q^a$ is a three-vector and $\gL^a_{\ b}$ is a $3\times 3$
Lorentz matrix. If we denote $V$ by $(\gL,q)$, then
Eq.\sx\ reads with $U^{-1}\equiv (\gL,q)$
$$ A_\mu = U\p_\mu U^{-1} \equiv \left(\matrix{\gL^{-1}\p_\mu\gL&
\gL^{-1}\p_\mu q\cr 0&0}\right) = \left(\matrix{\w^a_{\ \mu} J_a &
e^b_{\ \mu} P_b \cr 0&0}\right)
	\eqn\sxiii   $$
where
$$ (J_a)^b_{\ c} = \ge^{\ b}_{a\ c} \ , \qquad (P_a)^b = \gd^{\ b}_a
	\eqn\eq  $$
are the group generators. More explicitly
$$ J_0 = \left(\matrix{0&0&0\cr0&0&-1\cr0&1&0}\right) , \qquad
 J_1 = \left(\matrix{0&0&-1\cr0&0&0\cr-1&0&0}\right) , \qquad
 J_2 = \left(\matrix{0&1&0\cr1&0&0\cr0&0&0}\right)
	\eqn\eq  $$
Since the right-hand side of Eq.\sxiii\ is known,
we can solve the equation to obtain $(\gL, q)$.
To this end, we introduce
$$ J_+ \equiv {1\over 2}(J_0 + J_1) , \qquad
J_- \equiv {1\over 2}(J_0 - J_1)
	\eqn\eq  $$
Thus
$$ [J_2, J_+] = J_+ , \qquad [J_2, J_-] = -J_- , \qquad
[J_+, J_-] = -{1\over 2}J_2
	\eqn\sxvii  $$
We consider an arbitrary Lorentz group element $\gL$,
parametrized by\footnote{1}{The fact that any {\it rotation}
group element can be brought into
this form was used to construct representations for the three
dimensional Euclidean group, see \Ref\side{J.Elliott and P.Dawber,
{\it Symmetry in physics}, The Macmillan Press (1979).}.}
$$ \gL = e^{uJ_+-vJ_-}e^{wJ_2}
	\eqn\eq  $$
Using the commutators Eq.\sxvii\ and
\def\pa{\p_\ga}
\def\O{{\cal O}}
$$ \pa e^{\O} = \int^1_0 ds\ e^{s\O}\pa\O e^{(1-s)\O}
	\eqn\eq  $$
we compute \Ref\Witten{The calculation is similar to
that illustrated in C.Nappi and
E.Witten, Phys.Rev.Lett.{\bf 71}, 3751 (1993).}
$$\eqalign{&\qquad
\gL^{-1}\pa\gL = \pa wJ_2 + e^{-w}\pa uJ_+ - e^w\pa vJ_-
\cr & +  (u\pa v-v\pa u)\left[{1-\cosh\sqrt{uv}\over uv}J_2
+ {\sqrt{uv}-\sinh\sqrt{uv}\over (uv)^{3/2}}\left(ue^{-w}J_+
+ ve^wJ_-\right)\right]  }
	\eqn\eq   $$
Comparing this with Eqs.\sxiii, \conn\ and \sviii,
we get for type-I($+$) solutions
$$ \gL = e^{-\gb xJ_1}e^{SJ_2}
	\eqn\eq  $$
or, more explicitly,
$$ \gL = \left(\matrix{\cosh\gb x\cosh S&\cosh\gb x\sinh S&\sinh\gb x\cr
\sinh S&\cosh S &0\cr \sinh\gb x\cosh S&\sinh\gb x\sinh S&\cosh\gb x}
\right)
	\eqn\eq  $$
To find out the desired coordinate transformations, we only need to
integrate
$\p_\mu q = \gL e^b_{\ \mu} P_b$.
When this is done, we get $q = (\tt,\tz,\tx)^{\rm T}$ with
$$\eqalign{\tt &= {1\over \gb}B\cosh\gb x  \cr
\tx &= {1\over \gb}B\sinh\gb x  \cr
\tz &= \int dt\ A\sinh S = \int dz\ C\cosh S  }
	\eqn\eq  $$
where $S$ is given in terms of the metric via Eq.\sviii. Similarly, for
type-I($-$) solutions we have
$$ \gL = e^{\gb xJ_0}e^{SJ_2}
	\eqn\eq   $$
and the transformations are
$$\eqalign{\tt &= \int dt\ A\cosh S = \int dz\ C\sinh S  \cr
\tz &= {1\over \gb}B\cos\gb x  \cr
\tx &= {1\over \gb}B\sin\gb x   }
	\eqn\period  $$
For type-II($+$) solutions, the $\gL$ matrix is given by
$$ \gL = e^{x(J_0-J_1)}e^{(\ln P)J_2}
	\eqn\eq   $$
That is,
$$\gL = {1\over 2}\left(\matrix{(1+x^2)P + P^{-1}&
(1+x^2)P - P^{-1}&2x\cr
(1-x^2)P - P^{-1}&(1-x^2)P + P^{-1}&-2x\cr 2xP&2xP&2}\right)
	\eqn\eq  $$
The general transformations are
$$\eqalign{\tt &= {1\over 2}Bx^2 + {1\over 2}\int dt\ A\left(P +
{1\over P}\right) = {1\over 2}Bx^2 +
{1\over 2}\int dz\ C\left(P - {1\over P}\right)  \cr
\tz &= -{1\over 2}Bx^2 + {1\over 2}\int dt\ A\left(P -
{1\over P}\right) = -{1\over 2}Bx^2 +
{1\over 2}\int dz\ C\left(P + {1\over P}\right)  \cr
\tx &= Bx  }
	\eqn\eq  $$
where $P$ is given in terms of the metric via Eq.\six. Type-II($-$)
solutions are characterized by
$$ \gL = e^{-x(J_0+J_1)}e^{-(\ln P)J_2}
	\eqn\eq  $$
One can easily obtain the transformation laws.

	Next we specialize to solutions
with $A = C$.  We only consider ``$+$'' type solutions. The ``$-$''
type solutions can be similarly studied.
For type-II solutions, the integrability condition for
the first two equations in Eq.\six\
tells us that $\ln P$ satisfies the wave equation
$$ (\p_t^2 - \p_z^2)\ln P = 0
	\eqn\eq  $$
Thus the general solution for $P$ may be written as
$$ P = {f'^{1/2}(t+z)\over g'^{1/2}(t-z)}
	\eqn\eq  $$
where a prime denotes differentiation of a function with  respect to
its argument. The choice of this form for $P$ is for future convenience.
It follows that the metric is given by
$$ ds^2 = f'(t+z)g'(t-z)(-dt^2+dz^2) + f^2(t+z)dx^2
	\eqn\stiii  $$
where $f$ and $g$ are arbitrary functions subject to the constraint
Eq.\cons. Under a coordinate transformation
$$\eqalign{\tilde t + \tilde z &= f(t+z)  \cr
\tilde t - \tilde z &= g(t-z) + f(t+z)x^2  \cr
\tilde x &= f(t+z)x   }
	\eqn\tranone    $$
the above metric becomes Minkowski. Similarly, type-I, $A = C$
solutions are given by
$$ e^S = {f'^{1/2}(t+z)\over g'^{1/2}(t-z)}
	\eqn\eq  $$
and
$$ ds^2 = f'(t+z)g'(t-z)(-dt^2 + dz^2) + {1\over 4}\gb^2
[f(t+z) + g(t-z)]^2dx^2
	\eqn\stvii  $$
where $\gb$ is a constant and $f , g $ are
arbitrary functions satisfying
Eq.\cons. This can be transformed into Minkowski metric
by the transformations
$$\eqalign{ \tt &={1\over 2}[f(t+z)+g(t-z)]\cosh\gb x  \cr
\tx &= {1\over 2}[f(t+z)+g(t-z)]\sinh\gb x \cr
\tz &=  {1\over 2}[f(t+z)-g(t-z)]   }
	\eqn\trantwo  $$
Although the transformations Eqs.\tranone\ and \trantwo\ may be
directly obtained
without using the gauge formulation (They can certainly be verified
without reference to the gauge formulation.), the $x$ dependences in
these transformations are better understood within the gauge
formalism.

	Note that, the spacetime Eq.\stvii\ covers only
(a portion of) the region $|\tt| > |\tx|$, while there is no such
restriction for Eq.\stiii. (There could be other restrictions, of
course.) Spacetimes associated with Eq.\period\ are periodic
in $x$; other types of solutions do not share this property.

	We need to clarify what we mean by ``time-dependent
solutions''. Normally, when we say a spacetime is time-dependent, we
mean that it is locally time-dependent, that is,
it does not have a time-like Killing vector.
These vacuum spacetimes
are locally Minkowski and certainly not time-dependent. Once we
put a string at $z=0$, the spacetime will be globally
time-dependent if, in the Minkowski coordinates, the world sheet of the
string does not lie in a $\tt =$ constant surface (modulo a Lorentz
transform). Practically,
since all static solutions are obtained in \refmark{\Deser}, solutions
that are not simply related to those must be time-dependent ones.

	This completes our study of the vacuum solutions.

\chapter{ Open String Solutions and the Relation to Planar Walls }

	We will look for reflection symmetric solutions	of the form
$$ ds^2 = -A^2(t,|z|)dt^2 + C^2(t,|z|)dz^2 + B^2(t,|z|)dx^2
		\eqn\ssi  $$
The energy-momentum tensor of an open string, positioned at $z = 0$,
is given by
$$ T^{\mu\nu}(t,z) = S^{\mu\nu}(t)\delta(z)
	\eqn\eq   $$
We consider sources for which
$$ S^{\mu\nu}(t) = \tsigma(t)u^\mu u^\nu - \txi(t)(h^{\mu\nu} +
u^\mu u^\nu)
	\eqn\eq   $$
where $\tsigma $ and $\txi$ correspond to the energy density and tension
of the string, respectively. For the metric Eq.\ssi, the three velocity
of the string is $u^\mu = (1/A, 0, 0)$ and the normal to the $z = 0$
hypersurface is $n^\mu = (0, 1/C, 0)$. Thus the induced two dimensional
metric is
$$ h^{\mu\nu} = g^{\mu\nu} - n^\mu n^\nu = {\rm diag}
 \left(-{1\over A^2 }, 0 ,
{1\over B^2} \right)
	\eqn\eq  $$

	Conservation of the energy-momentum tensor is given by
$\nabla_\mu T^{\mu\nu} = 0$. The $x$ component of this equation
is trivially satisfied.
The $z$ component is also identically satisfied for reflection
symmetric solutions, due to the following prescription
\Ref\Guth{S.Blau, E.Guendelman and A.Guth,
Phys.Rev.D{\bf 35}, 1747 (1987).}
$$ \Gamma^z_{\alpha\alpha}(z = 0) = \lim_{\epsilon\rightarrow 0}
{1\over 2}\left[\Gamma^z_{\alpha\alpha}(z = +\epsilon) + \Gamma^z_{
\alpha\alpha}(z = -\epsilon)\right ] = 0
	\eqn\eq  $$
Finally, the $t$ component gives us the conservation law
$$ \pt(BC\tsigma) = \txi C\pt B
	\eqn\eq   $$

	In the regions $z > 0$ or $z < 0$, the
solutions are given by the
vacuum solutions found in the last section.
We need to match these solutions using junction
conditions. Junction conditions can be obtained by
integrating the Einstein equations
$$ R_{\mu\nu} - {1\over 2}g_{\mu\nu}R = 2\pi GT_{\mu\nu}
	\eqn\ssvii  $$
across the string. In an orthonormal basis, we have
$$\eqalign{R_{01} &= R_{0212} = 0  \cr
R_{22} &= -R_{0202} + R_{1212} = 2\pi G\tsigma\delta(z) \cr
R_{00} &= R_{0101} + R_{0202} = -2\pi G\txi\delta(z) \cr
R_{11} &= -R_{0101} + R_{1212} = 2\pi G(\tsigma + \txi)\delta(z) }
	\eqn\eq   $$
If we apply $\int^{+\epsilon}_{-\epsilon}dz$ to these equations,
only terms with
a $\delta(z)$ singularity survive. On the left-hand side,
only terms containing a double $z$ derivative $\p_z^2$ survive.
Using the explicit expressions
for the curvature tensor, Eq.\siv,
we obtain, for reflection symmetric solutions,
$$\eqalign{\tonee &= -\left.{1\over BC}\bc\right|_{z=+\epsilon}   \cr
\ttwoo &= -\left.{1\over AC}\ac\right|_{z=+\epsilon}    }
	\eqn\ssix  $$
For any solutions satisfying Einstein equations and the
junction conditions,
the energy-momentum conservation is automatically satisfied.

	One may compare these with the
junction conditions obtained using Gaussian normal
coordinates $(\tau , \eta , x)$ ,
in terms of which the metric is given by (For a discussion of
obtaining junction conditions with Gaussian normal coordinates,
see \refmark{\Guth} and
\Ref\normal{E.Farhi, A.Guth and J.Guven,
Nucl.Phys.B{\bf 339}, 417 (1990).}.)
$$ ds^2 = g_{\tau\tau}(\tau,\eta)d\tau^2 + d\eta^2 + \tilde B^2(\tau,
\eta)dx^2
	\eqn\eq  $$
where $g_{\tau\tau}(\tau, 0) = -1 , \tilde B(\tau,\eta) = B(t(\tau,\eta),
z(\tau,\eta))$ and $\eta(t, z=0) = 0 , z(\tau, \eta=0) = 0$.
The energy-momentum tensor
is $T^{\mu\nu}(\tau,\eta) = S^{\mu\nu}(\tau)\delta(\eta)$
with
$$S^{\mu\nu}(\tau) = \sigma(\tau)u^\mu u^\nu - \xi(\tau)(h^{\mu\nu}
+ u^\mu u^\nu)
	\eqn\ssxi  $$
Now $u^\mu = (1, 0, 0)$ and the two-dimensional induced metric is
$$ ds^2 = -d\tau^2 + \tilde B^2(\tau, 0)dx^2
	\eqn\eq   $$
The conservation law now reads
$$ \partial_\tau(\tilde B\sigma) = \xi\partial_\tau\tilde B
	\eqn\eq   $$
or equivalently, $\pt(B\sigma) = \xi\pt B$. The junction conditions
are
$$ K^i_j\Big|_{\eta=+\epsilon} = -\pi G(S^i_j-\delta^i_j{\rm Tr}S)
	\eqn\eq  $$
for reflection symmetric solutions. The non-zero components of
the extrinsic curvature are
$$\eqalign{K_{xx} &= {1\over 2}n^\mu\partial_\mu B^2 = {1\over C}
B\pz B  \cr
K_{\tau\tau} &= -n_\mu u^\nu\nabla_\nu u^\mu = -C\Gamma^z_{tt}u^tu^t
= -{\pz A\over AC}  }
		\eqn\eq   $$
So we get
$$\eqalign{\onee &= -\left.{1\over C}{\pz B\over B}
\right|_{z=+\epsilon}   \cr
\twoo &= -\left.{1\over C}{\pz A\over A}\right|_{z=+\epsilon}   }
	\eqn\ssxvi   $$
Eqs.\ssix\ and \ssxvi\ look different, but they
can be reconciled by noting that
$\delta(z)= C\delta(\eta)$ and hence that $\sigma = C\tsigma$ and
$\xi = C\txi$.

	We present some special solutions as examples. These
are obtained by choosing $S$ and $P$
in Eqs.\sviii\ and \six\ to be simple
functions. We use
Eq.\ssxi\ for the energy-momentum tensor, and thus
Eq.\ssxvi\ for the junction
conditions.  For type-I solutions we obtain
$$\eqalign{ds^2 &= -dt^2 + dz^2 + (1 - \alpha|z| + \beta t)^2dx^2  \cr
&\pi G\sigma = {\alpha\over 1 + \beta t}\ , \qquad \pi G\xi = 0 }
	\eqn\dust  $$
$$\eqalign{ds^2 &= -(1-\alpha|z|)^2dt^2 + dz^2 + (1-\alpha|z|)^2\sinh^2
\alpha t\ dx^2  \cr
&\pi G\sigma = \alpha , \qquad \pi G\xi = \alpha }
	\eqn\non   $$
and
$$\eqalign{ ds^2 &= -dt^2 + (1+\ga t)^2dz^2 + (1+\ga t)^2\cosh^2\ga z\
dx^2\cr
&\onee = \twoo = 0  }
	\eqn\vacuum  $$
where $\ga,\beta$ are arbitrary constants. They correspond to
choosing $S = {\rm constant}, -\ga t$ and $\ga z$, respectively.
Eq.\dust\ is a
generalization of the static dust string solutions \refmark{\Deser}.
Note that there are no static
reflection symmetric solutions with $\sigma$ and $\xi$
both non-vanishing.
This can be easily seen from Eq.\static\ and the junction conditions
Eq.\ssxvi. Eq.\vacuum\ describes a pure vacuum solution.
Type-II solutions can be similarly
constructed. We have
$$\eqalign{ds^2 &= -(1-\alpha|z|)^2dt^2 + dz^2 + (1-\alpha|z|)^2
e^{2\ga t} dx^2  \cr
&\pi G\sigma = \alpha , \qquad \pi G\xi = \alpha }
	\eqn\ssxx   $$
$$\eqalign{ds^2 &= -dt^2 + (1-\alpha t)^2dz^2 + (1-\alpha t)^2
e^{-2\ga|z|}dx^2  \cr
&\pi G\sigma = {\ga\over 1 + \ga t}\ , \qquad \pi G\xi = 0 }
	\eqn\eq   $$
where $\ga$ is a constant. They correspond to $P = \ga e^{\ga t},
\ga e^{-\ga z}$, respectively.

	One can transform these solutions into the $A = C$ form.
For example, from Eqs.\non\ and \ssxx\ we get
$$\eqalign{ds^2 &=e^{-2\ga|z|}(-dt^2+dz^2)+e^{-2\ga|z|}\sinh^2\ga t\
dx^2 \cr
&\onee=\twoo=\ga  }
	\eqn\eq   $$
and
$$\eqalign{ds^2 &=e^{-2\ga|z|}(-dt^2+dz^2)+e^{-2\ga|z|}e^{2\ga t}dx^2 \cr
&\onee=\twoo=\ga  }
	\eqn\eq   $$
respectively.
The coordinate singularities at $z = \pm 1/\ga$ in the original
solutions are now transformed to $z = \pm\infty$.

	Solutions with $\onee =\twoo =\ga = {\rm constant}$
are sometimes called vacuum string solutions.
They correspond to domain wall solutions in
$3+1$ dimensions. General vacuum string solutions with $A=C$ can be obtained
by solving the junction conditions Eq.\ssxvi.
Type-I solutions are found to be given by Eq.\stvii\ with
$$ g'^{1/2}=-{f'^{1/2}\over \ga f}\ ,\qquad
g=-{1\over \ga^2 f}
	\eqn\ssxxiv   $$
where $f$ is still an arbitrary function. Similarly, type-II solutions
are given by Eq.\stiii\ with $g$ given by Eq.\ssxxiv. One can transform
these into Minkowski metric using Eqs.\tranone\ and \trantwo. For both
types of solutions, the world sheet of the string is transformed into
(a portion of) the hyperboloid
$$ -\tt^2 + \tz^2 + \tx^2 = {1\over \ga^2}
	\eqn\eq   $$

	As point sources solutions in $2+1$ dimensions are related
to cosmic string solutions in $3+1$ dimensions, so should open string
solutions be related to planar wall solutions. Indeed, the solution
of Eq.\ssxx\ exhibits a remarkable resemblance to Vilenkin's domain
wall solution \refmark{\Vilenkin}, which takes the form
$$\eqalign{ds^2 &= -(1-\alpha|z|)^2dt^2 + dz^2 + (1-\alpha|z|)^2
e^{2\ga t} ( dx^2 + dy^2) \cr
&\pi G\sigma = 2\alpha , \qquad \pi G\xi = 2\alpha }
	\eqn\eq   $$
in terms of our convention. On the other hand, there is no
four-dimensional solution corresponding to Eq.\non; and one can verify
that
$$ds^2 = -(1-\ga z)^2dt^2 + dz^2 + (1-\ga z)^2\sinh^2\ga t\
(dx^2 + dy^2)
	\eqn\eq  $$
is not a vacuum solution in $3+1$ dimensions.

	To establish the general relations, we use
the reduction formulae
$$\eqalign{ds^2 &= g_{\ga\beta}dx^\ga dx^\beta + N^2(dx^3)^2   \cr
K_{\ga\beta} &= {1\over 2N}\p_3g_{\ga\beta}   \cr
{\mathop R^4 }_{\ga\beta\mu\nu} &= {\mathop R^3 }_{\ga\beta\mu\nu}
+ (K_{\beta\mu}K_{\ga\nu}-K_{\beta\nu}K_{\ga\mu})  \cr
{\mathop{ R^3}^4 }_{\beta\mu\nu} &= {1\over N}(D_\nu K_{\beta\mu}
- D_\nu K_{\beta\nu})   \cr
{\mathop{ R^3}^4 }_{\beta3\nu} &= -{1\over N}\p_3K_{\beta\nu}
-{1\over N}D_\nu D_\beta N + K_{\ga\beta}K^\ga_\nu    }
	\eqn\eq    $$
where $\ga,\beta,\mu,\nu = 0,1,2$ or $t,z,x$ and $D_\ga$ is the
covariant derivative associated with the three dimensional metric
$g_{\ga\gb}$
$$D_\beta T_\ga\equiv
\p_\beta T_\ga -{\mathop{ \gG^\gl}^3}_{\ga\beta} T_\gl
	\eqn\eq  $$
We see that given
a $2+1$ dimensional vacuum solution Eq.\si, the metric
$$ ds^2 = -A^2(t,z)dt^2 + C^2(t,z)dz^2 + B^2(t,z)(dx^2 + dy^2)
		\eqn\ssxxvii  $$
will be a flat
spacetime solution in $3+1$ dimensions if and only if
$$ D_\ga D_\beta B = \p_\ga\p_\beta B - \gG^\gl_{\ga\beta}\p_\gl B = 0
	\eqn\eq  $$
Performing a dimensional reduction once more to $1+1$ dimensions, we see
that only possibly non-vanishing terms are $D_\ga D_x B$. Using the
explicit
expressions for $\gG^\gl_{\ga\beta}$, we see that the only possible
non-zero term is
$$ -D_xD_xB = \gG^t_{xx}\p_tB + \gG^z_{xx}\p_zB = B\left[
\left(\ba\right)^2
- \left(\bc\right)^2\right]
	\eqn\eq  $$
Therefore a three-dimensional vacuum solution corresponds to a
four-dimensional flat spacetime solution if it is a type-II solution.
Type-I
solutions don't even correspond to vacuum solutions in four dimensions.
On the other hand, given any flat spacetime solution in
four dimensions of the form Eq.\ssxxvii,
Eq.\si\ is automatically a type-II
vacuum solution in three dimensions.

	Next we consider junction conditions.
If we normalize $G$ as in Eq.\ssvii, the junction conditions
in $3+1$ dimensions (with energy-momentum tensor given by Eq.\ssxi)
are
$$\eqalign{ {1\over 2}\onee
&= -\left.{1\over C}{\pz B\over B}\right|_{z=+\epsilon}   \cr
{1\over 2}\pi G(-\sigma+2\xi)
&= -\left.{1\over C}{\pz A\over A}\right|_{z=+\epsilon}   }
	\eqn\eq   $$
We see that a domain wall solution in $3+1$ dimensions corresponds to
a vacuum string solution in $2+1$ dimensions (Both terms mean
$\sigma = \xi = {\rm constant}$).
All class-I domain wall solutions of Ipser and Sikivie are flat
\refmark{\Ipser}; they correspond to the type-II vacuum string
solutions in three dimensions.
Their class-II solutions are unphysical
because of the curvature singularity.
Our type-I vacuum string solutions do not have any curvature
singularities, since the curvature identically vanishes away
from sources in three dimensions. Thus there are more solutions
in $2+1$ dimensions.

	Ipser and Sikivie have noted the ``gravitational
repulsiveness'' of a planar wall \refmark{\Ipser}.
Consider a test particle,
initially at rest, in the vicinity of a planar wall.
{}From the geodesic equation we obtain its initial acceleration in the $z$
direction:
$$ \left.{d^2z\over d\gl^2}\right|_{z=+\epsilon} =
-\gG^z_{tt}\left({dt\over d\gl}\right)^2
= -{A\pz A\over C^2}\left({dt\over d\gl}\right)^2
	\eqn\eq  $$
{}From the junction conditions we see that this is positive if
$-\sigma+2\xi$ is positive. The wall is then said to be repulsive.
Essentially the same calculation applied to the $2+1$ dimensional
theory tells  us that a string is gravitationally
repulsive if the string tension $\xi$ is positive.

\chapter{Conclusions}

	In conclusion, we have found general, line symmetric vacuum
solutions and we have matched them across the string using
the junction conditions. For any metric, we have given the corresponding
string energy density and tension. We have studied the gauge
formulation of these solutions and found their local Poincar\'e group
elements. The relation of these solutions to the four dimensional planar
walls is also obtained.

	Recently, Cangemi {\it et al.} have studied the gauge
formulation of the $2+1$ dimensional black hole
in anti-de Sitter space \Ref\Cangemi{D.Cangemi, M.Leblanc and
R.B.Mann, Phys.Rev.D{\bf 48}, 3606 (1993).}. It would be interesting
to study string solutions and their gauge formulation when
a cosmological constant is present.

\ack

	I thank Hsien-Chung Kao, Professor Kimyeong Lee and Professor
Erick Weinberg for helpful discussions. I also thank Professor Erick
Weinberg for valuable comments on the manuscript, Professor Roman
Jackiw for a critical reading of the paper.

\appendix

	In this appendix, we study geodesics in the background metric
$$ ds^2 = -z^2dt^2 + dz^2 +z^2e^{2t}dx^2 \ , \qquad
-1 < z < \infty
	\eqn\ai   $$
which is related to the solution Eq.\ssxx\ with $\ga = 1$
(in the $z > 0$ region) by a coordinate transformation. The position
of the string is now at $z = -1$ and the coordinate singularity
at $z = 0$.
Free motion of a test particle can be determined from the Lagrangian
$$L = -z^2\left({dt\over d\gl}\right)^2 + \left({dz\over d\gl}
\right)^2 + z^2e^{2t}\left({dx\over d\gl}\right)^2
	\eqn\eq   $$
Since there is only one cyclic coordinate, it is difficult to
solve the differential equations. Thus we will
use the Hamilton-Jacobi equation,
which turns out to be separable
$$ -{1\over z^2}\left({\p W\over\p t}\right)^2 + \left({\p W\over \p z}
\right)^2 + {1\over z^2}e^{-2t}\left({\p W\over\p x}\right)^2 = E
	\eqn\eq   $$
Letting  $W(t,z,x) = xp_x + W_1(t) + W_2(z)$, we get
$$ \left({dW_1\over dt}\right)^2 - e^{-2t}p^2_x = \eta , \qquad
\left({dW_2\over dz}\right)^2 - {\eta\over z^2} = E
	\eqn\eq  $$
Hence
$$ W(t,z,x) = xp_x + \int\sqrt{ E + {\eta\over z^2}}dz +
\int\sqrt{ \eta + e^{-2t}p^2_x}dt
	\eqn\eq   $$
where $p_x, E, \eta$ are arbitrary constants. The motion is described by
$$\eqalign{x_0 &= {\p W\over \p p_x} = x + \int{e^{-2t}p_x\over
\sqrt{\eta + e^{-2t}p^2_x }}dt   \cr
\zeta &= 2{\p W\over\p\eta} = \int{dz\over z^2\sqrt{E + z^{-2}\eta }}
+ \int{dt\over\sqrt{\eta + e^{-2t}p^2_x}}  \cr
\gl - \gl_0 &= {\p W\over\p E} = {1\over 2}\int{dz\over\sqrt{E +
z^{-2}\eta }}  }
	\eqn\eq   $$
where $p_x , \eta , E , x_0 , \zeta , \gl_0$ are constants. The integrals
can all be carried out. As a simple example, we consider the motion
with $\eta = 0$. The geodesic is
$$ t=\ln\left|{1-\ga\gl\over\gl}\right| , \qquad z = \gl , \qquad
x ={\gl\over 1-\ga\gl}
	\eqn\eq   $$
where $\ga$ is a constant. The metric Eq.\ai\ can be flattened by the
transformation
$$\tt+\tz=ze^t,\qquad \tt-\tz=-ze^{-t}+ze^tx^2,\qquad \tx=ze^tx
	\eqn\eq  $$
In terms of these new coordinates, the above geodesic becomes
manifestly a straight line
$$\tt+\tz=1-\ga\gl, \qquad \tt-\tz=0, \qquad \tx=\gl
	\eqn\eq   $$
On the other hand, an arbitrary straight line in $(\tt, \tz, \tx)$
coordinates is not
necessarily a geodesic, since it may be out of the region that the
spacetime of Eq.\ai\ covers.

\refout

\end